\documentstyle[11pt]{article}
 
\def\pa{\partial}
 \def\G{\Gamma}
\def\a{\alpha} 
\def\b{\beta} 
\def\d{\delta} 
\def\e{\epsilon}

\def\l{\lambda} 
\def\m{\mu} 
\def\n{\nu} 
 
\def\p{\pi} 
\def\v{\varphi}
 
\def\s{\sigma} 
\def\t{\tau}

\def\mn{{\mu\nu}}
\def\be{\begin{equation}}
\def\ee{\end{equation}}
\def\cA{{\cal A}}
\def\cB{{\cal B}}
\def\ou{{\bar u}}
    
\begin{document}

\begin{flushright}

BRX TH-440\\

\end{flushright}

\begin{center}
{\large\bf ``Good Propagation" Constraints 
on Dual Invariant Actions in Electrodynamics
and on Massless Fields}
\end{center}

\begin{center}
{ S. Deser$^{a}$\footnote{deser@binah.cc.brandeis.edu},
  J. McCarthy$^{b}$\footnote{Jim.McCarthy@dsto.defence.gov.au $\,\,\,$
  Present Address: DSTO (ITD), PO Box 1500, Salisbury SA 5108, Australia},
  \"{O}. Sar{\i}o\~{g}lu$^{a}$\footnote{sarioglu@binah.cc.brandeis.edu}}
\end{center}

\begin{center}{\sl
$^{a}$ Department of Physics, Brandeis University, Waltham, MA 02454-9110, USA\\
$^{b}$ Department of Physics and Mathematical Physics, University
       of Adelaide, SA 5005, Australia}
\end{center}

\begin{quotation}
We present some consequences of non-anomalous propagation requirements
on various massless fields. Among the models of nonlinear
electrodynamics we show that only Maxwell and Born--Infeld also obey
duality invariance. Separately we show that, for actions depending
only on the $F_\mn^2$ invariant, the permitted models have $L \sim
\sqrt{1 + F^2}$. We also characterize acceptable vector--scalar
systems. Finally we find that wide classes of gravity models share
with Einstein the null nature of their characteristic surfaces.
\end{quotation}

The nonlinear electrodynamics of Born--Infeld (BI)
\cite{BI} has been the subject of frequent revivals,
not least because it enjoys two quite separate
properties, shared with Maxwell theory. The first
is that its excitations propagate without the shocks
common to generic nonlinear models \cite{002,boi,BB}, the
second is duality invariance \cite{003}. Here,
we want to complete this subject in several ways. 
Our primary result will be that imposition of both
duality and good propagation singles out BI and 
Maxwell, without even requiring the solutions to 
reduce to Maxwell in the weak field limit (something
that is used to select BI in the derivations of 
each separate demand). We will then show separately that for
actions depending only on the Maxwell invariant 
$\a \equiv \frac{1}{2} F^2_\mn$ (rather than
on both $\a$ and the other invariant, $\b \equiv \frac{1}{8}
\e^{\mn\s\t} F_{\mn} F_{\s\t}$, allowed in D=4) 
correct propagation implies the form $L \sim \sqrt{1+\a}$, 
which shares a square root (but not a determinant) form with BI.
In this connection, we will 
also discuss the scalar analogs of BI and characterize 
systems involving both scalars and vectors. Criteria
for physical propagation of hyperbolic systems can be 
derived in many ways; in a separate work \cite{ozg} we will consider
the ``complete exceptional" (CE) approach \cite{lax}
to these questions, in more detail. 
Since this method was useful in some of our derivations
below, we will use ``CE" to mean propagation without
emerging shocks, with characteristic flow remaining
parallel along the waves (but waves need not travel 
with the same speed in all directions.)

In the following we will be concerned with systems of partial
differential equations which arise as the variational equations of a
relativistic Lagrangian theory.  Thus they will be quasilinear 
(linear in highest derivatives) and the coefficient functions will
not explicitly depend on the coordinates.  Without loss of generality,
they can be reduced to a set of equations first order in derivatives.
For definiteness then, with $u$ an $N$-vector of fields, $\cA$ an
$N$x$N$ matrix and $\cB$ an $N$-vector (both arbitrary (smooth enough) 
functions of $u$), the equations of interest are 
\be 
\cA^\mu(u)\partial_\mu u + \cB(u) =0 \, . \label{gen} 
\ee 
Of course, the theory of such equations in arbitrary dimensions is
quite difficult, but we will be interested in the evolution of the
spatial boundary of a wave propagating into some given vacuum state.
So, with $\ou$ some smooth (say at least $C^1$) solution, at some
initial time we have some spatial region outside of which the
``state'' is the ``vacuum solution'' $\ou$, and across the boundary
surface the full solution $u$ is continuous but its first
derivative may not be.  We now consider the evolution of this initial 
``wavefront''.   

One discussion of this situation has been developed
(\cite{boi} and references therein) which we will follow here.  Let the
hypersurface $S$, specified by 
\be
\varphi(x) = 0 \, , 
\ee 
denote the surface of evolution of the initial wavefront; i.e., the
initial wavefront is the spatial surface $\varphi(x,0) = 0$.
Since the field $u$ is continuous, only the normal derivative will be
allowed to be discontinuous.  Choosing a local coordinate system
$x^\mu = (\varphi ,\psi^i)$, we can define
the ``first order discontinuity'' in a given quantity $f$ to be
\be
\d_1 f \equiv \left[{\partial f\over {\partial \varphi}}\right] \, , 
\ee 
where  
\be
[X] \equiv X|_{\varphi = 0+} - X|_{\varphi = 0-}   
\ee
and we will sometimes write $[X] \equiv \d_0 X$.

 Taking the discontinuity of (\ref{gen}) we obtain ($\varphi_\mu \equiv
\partial_\mu \varphi)$
\be
\left(\cA^\mu \varphi_\mu\right) \d_1 u = 0 \, , \label{disgen}
\ee
where the matrix $\cA = \cA(\ou)$.  Hence, since $\d_1 u \neq 0$, we
see that $S$ must be a characteristic surface; i.e., the characteristic  
equation
\be 
H(x,\varphi_\mu) = {\rm det} \left(\cA^\mu \varphi_{\m} \right) = 0 \,  
\ee
must hold on $S$, where $H$ is homogeneous of order $N$ in $\varphi_\mu$.
The characteristic curves, which solve
$$
{d x^\mu \over {ds}} = {\partial H \over {\partial \v_{\mu}}} \, ,
$$
are clearly tangential to the characteristic surface.

From (\ref{disgen}), it further follows that $\d_1 u$ may be expanded
in a basis of ${\rm Ker}\left(\cA^\mu \varphi_\mu\right)$, the
coefficients of this expansion being called the ``coefficients of
discontinuity''.  In general, the coefficients of discontinuity evolve 
according to a nonlinear differential equation.  Following the discussion 
in \cite{boi}, based on \cite{CV}, it can be shown that when we ensure
that the characteristic curves do not intersect locally (and thus 
``shock'' singularities do not develop), the evolution
of the equations of discontinuity is correspondingly {\it linear}.  We
will impose this condition as our specification of ``nonanomalous
evolution''.  It is the CE condition mentioned earlier, and can be
imposed as the condition that, on the characteristic surface
$H = 0$, we have \cite{boi}
\be
\d_0 H = 0 \, . \label{CEdef}
\ee

We now apply these ideas to the systems of interest, without
however showing any details of the underlying CE derivation.
For general nonlinear gauge invariant actions in
$D=4$ that depend only on field strengths but not explicit
derivatives, the Lagrangian $L(\a ,\b )$ must obey the
two equations (subscripts indicate partial differentiation)
$$ - L_{\a} \, (4 \, L_{\a\a} - L_{\b\b}) \, + \, 2 \, \a \,
[L_{\a\a} \, L_{\b\b} - (L_{\a\b})^2] = 0 \;\; ,  \eqno{(7a)} $$
$$ - L_{\a} \, L_{\a\b} \, + \, \b \, 
[L_{\a\a} \, L_{\b\b} - (L_{\a\b})^2] = 0 \;\; .  \eqno{(7b)} $$
in order for its excitations to be CE and for light to travel
according to only {\it one} dispersion law, i.e. no birefringence
\cite{boi,BB}.  To obtain them, however,
it was necessary to assume nontrivial $\b$ dependence;
see below. Quite independently the duality invariance 
requirement is given by \cite{003}
\setcounter{equation}{7}
\be
(L_{\a})^2 - \frac{\a}{2 \b} L_{\a} L_{\b} 
- \frac{1}{4} (L_{\b})^2 = \frac{1}{4} \;\; . \label{dual}
\ee
The simultaneous solution of $(7)$ and (\ref{dual}) 
can be obtained as follows. 

We first change to variables $(a,b)$ with 
$\a=a+b$, $\b^2=-a b$ (these new variables have the 
fundamental advantage that they factorize $L_{BI}$);
then $(7a)$, $(7b)$ and (\ref{dual}) transform into
$$ -4ab(a+b) (a-b)^2 [L_{aa}L_{bb}-(L_{ab})^2] +4 a b(a^2-b^2)L_{ab}(L_a-L_b)
+8 a b(b-a)L_{ab}(b L_b-a L_a)  \nonumber $$ 
$$ -2(a-b)^2 (a+b) [a L_b L_{aa}+b L_a L_{bb}] 
+ (4 a b+(a+b)^2)(L_a-L_b)(a L_a - b L_b) = 0, \eqno{(9a)} $$
$$ 2 a b(a-b)^2 [L_{aa}L_{bb}-(L_{ab})^2] +(a+b)(L_a-L_b)(b L_b-a L_a)+
(a-b)^2 [a L_b L_{aa}+b L_a L_{bb}] \nonumber $$
$$ -(a^2-b^2)L_{ab}(a L_a - b L_b) 
+2 a b (b-a)L_{ab}(L_a-L_b) = 0,
\eqno{(9b)} $$
\setcounter{equation}{9}
\be
L_a L_b = \frac{1}{4} \;\; , \label{newdu} 
\ee
respectively. Note the symmetry in $(a,b)$ of each equation. 
Multiplying $(9b)$ by $2(a+b)$ and adding the
result to $(9a)$, one finds
\be
(a-b)^2 (a L_a -b L_b) [2(b-a)L_{ab}-(L_a-L_b)] = 0 \;\;. \label{bir}
\ee
Vanishing of the $ (a L_a -b L_b) $ factor is not a useful solution
of (\ref{bir}) because it would imply \( L=L(ab) \equiv L(\b^2) \),
for which (\ref{dual}) implies 
$L \sim \b$, a total divergence.  Hence we must impose
\be
2 (b-a) L_{ab} - (L_a - L_b) = 0 \;\;. \label{new1}
\ee
Substituting this back into $(9a)$ gives the simplified form
\be
2 a b [L_{aa}L_{bb}-(L_{ab})^2] + 
[a {(L_a L_b)}_{a} + b {(L_a L_b)}_{b}] = 0 \;\;. \label{new2}
\ee
So now instead of (9) and (\ref{newdu}),
we can study (\ref{new1}), (\ref{new2}) and (\ref{newdu}). 
But actually it is easy to see that (\ref{new2}) follows
from (\ref{newdu}), since (\ref{newdu}) is a first integral
of the Monge--Amp\`{e}re factor \( [L_{aa}L_{bb}-(L_{ab})^2] \),
i.e. this factor vanishes as a consequence 
of (\ref{newdu}). Thus we are left with the system 
(\ref{newdu}), (\ref{new1}) \footnote{An alternative derivation 
inserts the general parametric solution 
\cite{ch} of (\ref{newdu}) into (\ref{new1}).}.

Substituting for $L_b$ in (\ref{new1}), it can be
written as \( (a-b)^2 \left( \frac{4(L_a)^2}{a-b} \right)_{b} =1 \),
with first integral
\be
L_a = \frac{1}{2} \sqrt{1+(a-b)f(a)}  \;\;. \label{lofa}
\ee
But by (\ref{newdu}), \( 4(L_a)^2 = L_a/L_b = 1+(a-b)f(a) \)
which gives \( L_b = \frac{1}{2} [1+(a-b)f(a)]^{-1/2} \), whose
integral is 
$$ L = - \frac{1}{f(a)} \sqrt{1+(a-b)f(a)} + h(a) \;\;. \eqno{(15a)} $$
Given the $(a,b)$ symmetry of the equations, the above procedure 
based on $L_b$ instead of $L_a$ gives the corresponding form
$$ L = - \frac{1}{k(b)} \sqrt{1+(b-a)k(b)} + m(b) \;\;. \eqno{(15b)} $$
Consistency with (\ref{newdu}) demands that
\( \sqrt{1+(a-b)f(a)} \sqrt{1+(b-a)k(b)} = 1 \), which implies
\( k(b)= \frac{f(a)}{1+(a-b)f(a)} \).
Substituting for $k(b)$ in $(15b)$ then shows that
\setcounter{equation}{15}
\be
L = - \frac{1}{f(a)} \sqrt{1+(a-b)f(a)} + m(b)  \label{lwfm}
\ee
which implies $h(a)=m(b)=d=const$. Differentiating (\ref{lwfm})
with respect to $a$ and comparing the result with (\ref{lofa})
one finally finds that \( (2+(a-b)f(a)) (f^{\prime} - f^2) = 0 \),
which when integrated gives (for $s$ an integration
constant) $f(a)=\frac{1}{s-a}$ (or $f=0$ trivially,
the Maxwell case.) Finally then, renaming constants, 
\( L= - \frac{1}{c} \sqrt{(1+ac)(1+bc)} + d \) (and 
$L=-\frac{1}{2} (a+b)$ for $f=0$). Rewriting these using 
$(\a,\b)$, we find using allowed rescalings that $L=-\frac{1}{2} \a$, 
Maxwell, and $L=1-\sqrt{1+\a-\b^2}=1-\sqrt{-\det[\eta_\mn+F_\mn]}$, 
Born-Infeld, are the only possible solutions. This demonstrates 
the uniqueness of BI and Maxwell as the simultaneously CE and 
duality invariant electrodynamics. The CE requirement (7) alone
permits additional solutions, such as $L=\a / \b$, that are not
duality invariant. [Conceivably, requiring power series 
expandability in $(\a,\b)$ might restrict the solutions of (7) to
Maxwell and BI.]

As was mentioned earlier, the CE criteria (7)
are only correct for nontrivial $\b$-dependence; for pure
$L(\a )$ they just imply $L^\prime= const.$, namely
Maxwell.  Instead, if one goes back to the complete CE 
requirements, they imply, for pure $L(\a)$ 
\be
L^\prime L^{\prime\prime\prime} - 3 (L^{\prime\prime})^2 = 0\; . \label{sca}
\ee
The solution, apart from Maxwell $(L^{\prime\prime}=0)$
is
\be
L (\a ) = k + (d + c \a )^{\frac{1}{2}} \; .
\ee
[We remark that in $D=3$, where $\a$ is the only invariant,
this is also the CE result, there also 
\( \sqrt{1+\a} = \sqrt{-\det[\eta_\mn+F_\mn]} \) with the 
BI determinant form. In $D=2$ there is of course no
propagation for any $L(\a)$ and correspondingly no 
restrictions are imposed.]  The above BI form is very
analogous to that obtained for a scalar field 
\cite{boi}.  There, (in any dimension) for $L(z)$, 
$z \equiv \frac{1}{2} (\pa_\m\phi) (\pa^\m \phi)$ being the only invariant
(in first derivatives), we find the same requirement (\ref{sca}). 
Hence we obtain, apart from the free scalar solution $L=-\frac{1}{2} z$,
 the same square root solution. Amusingly, this one can be put into 
BI-like form, since (rescaling $\phi$)
\be
\sqrt{1+2z} \equiv \sqrt{1+\phi^2_\m}
= \sqrt{-\det [\eta_\mn + \phi_\m\phi_\n]}
\ee
where $\phi_\m \equiv \pa_\m \phi$ denotes the field strength.
If one combines Maxwell and the (neutral) scalar into an
action $L(\a ,\b , z)$, then the CE conditions  further 
require $L_{z\a} = 0 = L_{z\b}$, reducing the Lagrangian
to the noninteracting $L(\a ,\b ) + L(z)$ form.  
Having the ``fully" BI form
$\sqrt{-\det [\eta_\mn + F_\mn + \phi_\m\phi_\n ]}$ in mind,
one can actually show more generally that there are no CE
actions with nontrivial dependence on the other possible
variable $y \equiv \frac{1}{2} (F_\mn \phi^\n)^2$.

Finally, we turn to gravitation. For Einstein's gravity in vacuum, 
as well as the linearized theory, the gravitational waves are CE, 
the characteristic surfaces describing discontinuities being null
(see e.g. \cite{lic}). It can be shown that this result holds for any $D>4$. 
[For $D=3$, there is of course no propagation and no restrictions
are imposed.] One can further look at pure gravitational actions 
of the form $p R^2_\mn - q R^2$ in $D=4$ and $f(R)$ in $D>3$ and 
show that the same conclusion remains unchanged.

To reduce these theories to a first order system would be
inconvenient, but is fortunately made unnecessary by a simple
extension of the previous discussion. Clearly, if we rebuilt
the original higher order equations from the set (\ref{gen}),
we would simply have the situation that all the derivatives
of the field are assumed continuous except the highest one. Thus
for quasilinear systems higher order, say $q$, in derivatives,
we define
\[ \d_r f \equiv \left[ \frac{\partial^{r} f}
{\partial \v^{r}} \right] \, , \]
and will consider the case that
\[ \d_q u \neq 0 \, ; \;\;\; \d_r u = 0 \; , 0 \leq r < q \; . \]
Notice that 
\[ \d_r \partial_{\m} = \partial_{\m} \v \, \d_{r+1} \; . \]

Let us first sketch the Einstein case to establish notation.
Considering a second order discontinuity in the metric across
some characteristic surface $\v=0$, $\d_2 g_{\mn}=\p_{\mn}$,
we have $( \v_{\m} \equiv \partial_{\m} \v )$
\[ \d_1 \G^{\l}~_{\mn} = \frac{1}{2} 
( \v_{\m} \p^{\l}~_{\n} + \v_{\n} \p^{\l}~_{\m}
- \v^{\l} \p_{\mn} ) \;\;, \] 
\begin{eqnarray*}
\d_0 R_{\mn} & = & \v_{\l} (\d_1 \G^{\l}~_{\mn}) -
                   \v_{\n} (\d_1 \G^{\l}~_{\l\m}) \\
             & = & \frac{1}{2} ( \v_{\m} \v_{\l} \p^{\l}~_{\n} +
\v_{\n} \v_{\l} \p^{\l}~_{\m} - \v_{\m} \v_{\n} \p^{\l}~_{\l} -
\v_{\l} \v^{\l} \p_{\mn} )                   
\end{eqnarray*}
and 
\[ \d_0 R = g^{\mn} (\d_0 R_{\mn}) = 
\v^{\m} \v^{\n} \p_{\mn} - \v_{\m} \v^{\m} \p^{\n}~_{\n} \]
which implies for 
\[ \d_0 G_{\mn} = \d_0 (R_{\mn} - \frac{1}{2} g_{\mn} R) =
   \d_0 R_{\mn} - \frac{1}{2} g_{\mn} \d_0 R = 0  \] 
\be 
\d_0 G_{\mn} = \frac{1}{2} \left[ \v_{\m} \v_{\l} \p^{\l}~_{\n} +
\v_{\n} \v_{\l} \p^{\l}~_{\m} - \v_{\m} \v_{\n} \p^{\l}~_{\l} -
\v_{\l} \v^{\l} \p_{\mn} - g_{\mn} 
(\v^{\s} \v^{\t} \p_{\s\t} - \v_{\s} \v^{\s} \p^{\t}~_{\t}) 
\right] = 0 \;\;. \label{ein}
\ee

In the harmonic gauge \( g^{\mn} \G^{\s}~_{\mn} = 0 \), one finds that
its first discontinuity implies
\be
2 \p^{\mn} \v_{\m} - \p^{\m}~_{\m} \v^{\n} = 0  \label{gbir}
\ee
Multiplying this by \( g_{\n\s} \v_{\t} + g_{\n\t} \v_{\s} \),
one gets
\be
\v_{\m} \v_{\l} \p^{\l}~_{\n} + \v_{\n} \v_{\l} \p^{\l}~_{\m} - 
\v_{\m} \v_{\n} \p^{\l}~_{\l} = 0 \;\;,  \label{giki}
\ee
whereas contracting by $\v_{\n}$, one finds
\be
\v^{\m} \v^{\n} \p_{\mn} = \frac{1}{2} \v_{\m} \v^{\m} \p^{\n}~_{\n}
\;\; .  \label{guc} 
\ee
Using (\ref{giki}) and (\ref{guc}) in (\ref{ein}), one ends up with
\[ \d_0 G_{\mn} = \frac{1}{2} ( \v_{\l} \v^{\l} \p^{\mn}
+ \frac{1}{2} g_{\mn} \v_{\l} \v^{\l} \p^{\s}~_{\s} ) = 0 \;\; . \]
Hence taking the trace
\[ \d_0 G^{\m}~_{\m} = \frac{(D+2)}{4} 
\v_{\l} \v^{\l} \p^{\s}~_{\s} = 0 \;\; . \]
The discontinuity in $g_{\mn}$ is arbitrary, hence 
$ \p^{\s}~_{\s} \neq 0 $, which implies that
$\v_{\l} \v^{\l}=0$. This tells that the characteristic surfaces are
null: the discontinuities travel with the speed of light in all
directions. The same holds for the linearized version of the theory
as well of course.

For generic quadratic Lagrangians $ (p R_{\mn} R^{\mn} - q R^2) 
\sqrt{-g}$ in $D=4$, using similar steps (writing the field 
equations, choosing harmonic gauge as before and utilizing the identities 
(\ref{giki}), (\ref{guc})) one finds that $( Q \equiv \v^{\l} \v_{\l} \;,
\p \equiv \p^{\l}~_{\l} )$
\be
Q \left( \frac{1}{2} (p-2q) \v_{\m} \v_{\n} \p -
\frac{p}{2} Q \p_{\mn} - \frac{1}{2} g_{\mn} (\frac{p}{2}-2q) Q \p 
\right) = 0 \;\; . \label{ste}
\ee
Taking the trace, one gets $Q^2 \p (p-3q) = 0$. (The choice $p=3q$
corresponds to Weyl--tensor squared; the scalar degree of freedom
is absent.) For $p=3q$, (\ref{ste}) becomes
\[ q Q ( \frac{1}{2} \v_{\m} \v_{\n} \p - \frac{3}{2} Q \p_{\mn}
+ \frac{1}{4} g_{\mn} Q \p ) = 0 \;\; . \]
Since $\p_{\mn}$ is arbitrary, we see that again $Q=0$, as in Einstein,
so $Q=0$ characterizes both Einstein and the quadratic action.

For actions $f(R) \sqrt{-g}$ in $D \geq 4$, the field equations are
\[ E_{\mn} \equiv R_{\mn} f^\prime - \frac{1}{2} g_{\mn} f +
(g_{\mn} \nabla_{\s} \nabla^{\s} - \nabla_{\m} \nabla_{\n})
f^\prime = 0 \;\; . \]
Hence the order of highest derivatives is four. Following similar
steps by taking $\d_4 g_{\mn}=\p_{\mn}$, we find the same 
expressions for $\d_3 \G^{\l}~_{\mn}$ and $\d_2 R_{\mn}$ as for 
$\d_1 \G^{\l}~_{\mn}$ and $\d_0 R_{\mn}$ in the Einstein case.
Using these, we get
\[ \d_0 E_{\mn} = (Q g_{\mn} - \v_{\m} \v_{\n})
(\v^{\s} \v^{\t} \p_{\s\t} - Q \p ) f^{\prime\prime} = 0 \;\; . \]
Finally, in the harmonic gauge with identity (\ref{guc}) and
taking the trace, one gets
\[ \d_0 E^{\m}~_{\m} = \frac{(1-D)}{2} Q^2 \p f^{\prime\prime} = 0 \; . \] 
Here too $Q=0$ is the only solution, and so for a wide class of
gravitational actions the propagation obeys the Einstein behavior.

This work was supported in part by NSF, under grant no 
PHY-9315811.

\end{document}